 \documentstyle[aps,epsf]{revtex}
 
\begin{document} 
 \title{Shrinking II --- The Distortion of the Area Distance-Redshift 
Relation in Inhomogeneous Isotropic Universes} 
 \author{ Nazeem Mustapha,\footnote 
             {email:~nazeem@maths.uct.ac.za} 
          B.A. Bassett,\footnote 
             {email:~bruce@stardust.sissa.it} 
          Charles Hellaby,\footnote 
             {email:~cwh@maths.uct.ac.za} 
          and  G.F.R. Ellis\footnote 
             {email:~ellis@maths.uct.ac.za}} 
 \address{Department of Mathematics and Applied Mathematics, 
University of Cape Town, Rondebosch, 7700, South Africa} 

 \maketitle 

 \begin{abstract}                                            
 {\bf Abstract.} This paper and the others in the series 
\cite{bi:EBD96,bi:SEB96} challenge the standard model of the effects 
of gravitational lensing on observations at large distances. We show 
that due to the cumulative effect of lensing, areas corresponding to 
an observed solid angle can be quite different than would be estimated 
from the corresponding Friedmann-Lema\^{\i}tre model, even when 
averaged over large angular scales. This paper concentrates on the 
specific example of spherically symmetric but spatially inhomogeneous 
dust universes, the Lema\^{\i}tre-Tolman-Bondi models, and shows that 
radial lensing significantly distorts the area distance-redshift and 
density-redshift relations in these exact solutions compared with the 
standard ones for Friedmann-Lema\^{\i}tre models.  
 Thus inhomogeneity may introduce significant errors into distance 
estimates based on the standard {\sc fl} relations, even after 
all-sky averaging. In addition a useful new gauge choice is presented 
for these models,  solving the problem of locating the past null cone 
exactly. 
\end{abstract} 

 \draft 
 \pacs{04.20.Cv, 95.30.Sf, 98.60.Eg, 98.60.Gi, 98.80.Es} 

 \begin{center} 
 Short Title:~~~~Shrinking II \\[2mm]
 Submitted to Class. Q. Grav.~~ March 1997 
 \end{center} 

 \section{Introduction and General Arguments} 

Our aim in this series of papers is to show that when analyzing 
observations at large redshift in the real universe, the assumption of 
an area distance corresponding to that of a 
 best-fit 
 Friedmann-Lema\^{\i}tre ({\sc fl}) model\footnote 
 {We follow recent moves to use `Friedmann-Lema\^{\i}tre' to describe 
the dynamics of and the application of the Einstein field equations to 
the standard model, and `Robertson-Walker' to describe its metric and 
geometry.} 
 --- that is, one with a matching averaged matter density 
 --- may not be a good approximation, even when averaging over 
large angular scales.  This claim is important because of the 
ubiquitous use of {\sc fl} models in studies of number counts versus 
redshift and area distance versus redshift. 

The first paper in the series (paper I) \cite{bi:EBD96} gave general 
arguments for this thesis. It was explained there that whilst the 
Dyer-Roeder distance \cite{bi:DR72} is generally regarded as a good 
approximation for ray bundles moving between high-density clusters of 
matter, resulting in a de-focussing relative to the comparable {\sc 
fl} model, matter moving near or through higher density regions is 
more focused than in the {\sc fl} model, so resulting in a 
compensating effect. It is commonly believed \cite{bi:W76,bi:SEF92} 
that an exact cancellation between the two effects takes place when 
one averages over large angular scales containing both high and low 
density regions 
 --- so the {\sc fl} area distance is the correct one on these  
scales. However, it was pointed out in \cite{bi:EBD96} that, after 
caustics have formed  through the focussing of light rays by the high 
density regions, these light rays too are rapidly diverging so that 
sufficiently far down the past light cone {\it all} light rays will be 
diverging relative to the comparable {\sc fl} model. 
 Consequently there is good reason to question the general opinion in 
this matter 
 --- we find that in fact `shrinking' takes place.\footnote 
 {The term shrinking \cite{bi:L88} describes the following.  If at 
some redshift the area distance is greater in the inhomogeneous 
universe than in the best-fit or background {\sc fl} model, scales at 
that redshift will be underestimated if the {\sc fl} area distance is 
used to calculate the size of objects from their measured angular 
size, and so will have `shrunk' relative to their actual scales in the 
more realistic model.  Put another way, if a sphere at given $z$ about 
the observer has a larger total area in the inhomogeneous model, an 
object of fixed size will subtend a smaller solid angle on the sky, 
and so be shrunk.  (See \cite{bi:EBD96} for more discussion.)} 
  It is often argued that photon conservation excludes this 
possibility, but these arguments are based on incorrect assumptions 
about the geometry. 

Given this feature, even when there is only weak lensing,\footnote 
 {We refer to `lensing' when the light rays are different than they 
would be in the background {\sc fl} model. Thus weak lensing implies a 
change of apparent positions and alteration in the usual distance 
relations, usually combined with image distortions, but not 
necessarily multiple imaging.} 
 and caustics have not formed, it is not obvious that the {\sc fl} 
area distance-redshift relation is an accurate description. 
 
 The aim of this paper is to construct an exact inhomogeneous model 
and its {\sc fl} approximation, and compare the area distance-redshift 
and density-redshift relations in the two. To do so we examine 
Lema\^{\i}tre-Tolman-Bondi ({\sc ltb}) spherically symmetric dust 
solutions \cite{bi:L33,bi:T34,bi:B47}, where exact integrations of the 
field equations are available for the past light cone of observers at 
the central position.   Although the lensing that occurs for this 
central observer is purely radial, we find the inhomogeneity has a 
tangible effect on observational relations.\footnote 
 {Radial lensing is a spherically symmetric distortion of the null 
cone compared with an {\sc fl} model, resulting in a uniform delay of 
the wavefront.  There is no image distortion, no dependence of 
magnification or time delay on direction, and no multiple imaging, but 
our results show its effects are observable.} 
  In the particular case examined, because of the spherical symmetry 
about the observer, the effect in any specific direction will not be 
compensated by an opposite effect in another direction 
 --- on the contrary, the effect is uniform because it is the same in 
all directions, and will occur on large as well as small angular 
scales. However unlike the previous paper \cite{bi:EBD96}, this will 
not be associated with the formation of caustics. 

  The failure of the {\sc fl}-like area distance-redshift assumption  
will thus have been shown to occur for observers at a very special 
position in a family of high-symmetry space-times. Undoubtedly this is 
not the kind of situation the authors of the papers mentioned above, 
claiming the effect does not occur, had in mind. However the usual 
statements of this effect contain no clauses that exclude this 
situation: the result is supposed to hold in all cases when the 
`lumpy' universe is reasonably close to an {\sc fl} model, and 
observations are averaged over the sky; their arguments do not exclude 
the situation envisaged here where the observer is at the centre of a 
spherically symmetric inhomogeneity. Indeed, a spherically symmetric 
model may be regarded as describing data that has been averaged over 
the {\em whole} sky, but not over distance. 

Our example thus confirms the claims of paper I, in the setting of 
particular exact inhomogeneous solutions of the Einstein Field 
Equations. It does not generically establish the magnitude of the 
effect, precisely because the high-symmetry geometry considered here 
precludes formation of caustics and the consequent fractal-like 
structure of the real light cone. Paper III \cite{bi:SEB96} will 
provide confirmation of the overall shrinking effect due to caustics, 
and attempt more realistic estimates of its magnitude than those given 
in paper I, which used simple analytic formulae for this purpose. 

In developing the results of this paper, we solve one of the problems 
that has made analysis of observations in  Lema\^{\i}tre-Tolman-Bondi 
solutions difficult, namely the problem of precisely locating the past 
light cone of the chosen central event P, by use of a special choice 
of radial coordinate that ensures a very simple form for the past 
light cone of P in these inhomogeneous space-times. This technical 
development has other uses in terms of analysing observational 
relations in these models.\footnote 
 {For a slightly different analysis of {\sc ltb} spacetimes, based on 
null cone coordinates, see \cite{bi:MHMS95}, and for a consideration 
of observations away from the centre of symmetry see \cite{bi:HMM96}.} 
 Our approach is complementary to that of Kurki-Suonio and Liang 
\cite{bi:K-SL92}, who did numerical calculations of observational 
relations in a hyperbolic {\sc ltb} model derived from an $\Omega = 
0.1$ {\sc fl} model plus some overdensities. 

 \section{Programme} 
 \label{sec:prog} 

We select the simplest inhomogeneous fluid solution of the Einstein 
Field Equations; the Lema\^{\i}tre-Tolman-Bondi ({\sc ltb}) model 
which is spherically symmetric, but radially inhomogeneous, with a 
dust equation of state. 

The question we are raising is whether the area of an averaged 
wavefront we receive at our observatory in an inhomogeneous universe 
is the same as the area of a wavefront in a smoothed version of that 
universe.  To clarify this  issue our strategy is to: 
 \\ $\bullet$ Select the most natural generalisation of the Einstein-
de Sitter models commonly used in studies of observations (i.e. a 
parabolic {\sc ltb} model) and describe data on the null cone. 
 \\ $\bullet$ Find the {\sc fl} limit of this inhomogeneous universe 
in an appropriate coordinate system. 
 \\ $\bullet$ Average the lumpy universe in a natural way and fit it 
correctly to a {\sc fl} model. 
 \\ $\bullet$ Compare area distances in the lumpy universe and its 
smoothed average. 

 \subsection{The Inhomogeneous Model} 
 \subsubsection{The Integrated Field Equations} 
 We choose the parabolic {\sc ltb}  model which is the natural  
generalisation of the $\Omega = 1$ dust {\sc fl} model. This model is 
characterised by the mass within comoving radius $r$, $M(r)$, and the 
`bang-time' function $t_B(r)$ describing the locus of the initial 
spatial hypersurface (that is, the local time of the big bang).  The 
two arbitrary functions represent a physical freedom and a coordinate 
freedom, e.g. $t_B(M)$ and $M(r)$, respectively. 

In normalised comoving coordinates the metric after solving the 
 off-diagonal EFE is 
 \begin{equation} 
 ds^2 = -dt^2 + (R')^2 dr^2 + R^2(d\theta^2 + \sin^2 \theta d\Phi^2) 
 \label{eq:flatmet} 
 \end{equation} 
 where $R'(t,r) = \partial R(t,r)/\partial r$. 

The time curves are irrotational and, for comoving dust ($p = 0)$, are 
necessarily geodesics because of momentum conservation. The spatial 
sections are flat because if we choose $r = R(t_0,r)$ then $R'(t_0,r) 
= 1$ and we find that the 3-spaces have metric $d\sigma^2 = dr^2 + r^2 
d\Omega^2$ and hence are flat. 

The areal radius, $R = R(t,r)$ in the Lema\^{\i}tre-Tolman-Bondi  
metric, {\it is} the area of the intersection of our past null cone 
with past spacelike time surfaces (in this case spheres). In the 
parabolic case $R$ is given explicitly by the solution to the equation 
of motion 
 \begin{equation} 
 {\dot{R}(t,r)} = \sqrt{\frac{M(r)}{R(t,r)}} 
 \label{eq:Rdot} 
 \end{equation} 
 obtained from the {\small{11}}, {\small{22}} and {\small{33}} 
components of the EFE, where $\dot{{}}$ denotes the derivative with 
respect to $t$; i.e. 
 \begin{equation} 
 R(t,r) = \left[ \frac{9 M(r)}{2} (t - t_{B}(r))^2 \right]^{1/3} 
 \label{eq:R} 
 \end{equation} 
 where $t$ is cosmic time whilst $M$ and $t_B$ are both functions of 
coordinate radius $r$ only. It follows immediately that 
 \begin{equation} 
 R'(t,r) = {3 \over 2 R^2(t,r)} 
 \left[ M'(r) (t - t_B(r))^2 - 2 M(r) (t - t_B(r)) t'_B(r) \right] 
 \label{eq:rprime} 
 \end{equation} 
 where $'$ denotes the derivative with respect to radial coordinate 
$r$. 

The {\small{00}} field equation gives                
 \begin{equation} 
 4\pi \rho(t,r) = \frac{M'(r)}{R^2(t,r) R'(t,r)}\,. 
 \label{eq:rho} 
 \end{equation} 

 \subsubsection{The Solution on the Null Cone} 
 Since we are interested in observations on the null cone we must  
project onto it by specifying the unique relationship between $r$ and 
$t$. On radial null geodesics, $ds^2 = 0 = {d\theta}^2= {d\Phi}^2$, so 
from (\ref{eq:flatmet}), if the past light cone of the event $(t = 
t_0, r = 0)$ is given by $t = \hat{t}(r)$, then that light cone is 
described by 
 \begin{equation} 
 dt = \pm R'(\hat{t}(r),r) dr\,. 
 \end{equation} 
 The coordinate freedom in the {\sc ltb} metric is a rescaling of the 
radial coordinate $r \rightarrow \tilde{r} = \tilde{r}(r)$. If we 
choose $r$ so that 
 \begin{equation} 
 R'(\hat{t}(r),r) = 1 \,, 
 \label{eq:ref1} 
 \end{equation} 
 then on the past light cone $dt = -dr$, so that the incoming light 
rays at the event $(t = t_0, r = 0)$ are given by 
 \begin{equation} 
 \hat{t}(r) = t_0 - r \,. 
 \label{eq:ref3} 
 \end{equation} 
 So this gauge choice, in contrast to other work done on observations 
in the {\sc ltb} model, locates  the null cone of the observer at one 
instant of time, in its simplest possible form, and makes our 
programme analytically solvable. On this light cone, putting 
(\ref{eq:ref3}) in (\ref{eq:rprime}) and using (\ref{eq:ref1}),  
 \begin{equation} 
 R'(\hat{t}(r),r) = {3 \over 2 R^2(\hat{t}(r),r)} [M'(r) \tau(r)^2 + 2 
M(r) \tau(r) \tau'(r) + 2 M(r) \tau(r)] = 1 
 \label{eq:condn} 
 \end{equation} 
 where we have defined 
 \begin{equation} 
 \tau(r) = t_0 - r - t_B(r)\,.  
 \end{equation} 
 The function $\tau(r)$ can be interpreted as proper time from the 
bang surface to our past null cone along the particle worldlines. We 
can set $t_0$ to be the time since the bang at the observer ($r = 0$) 
by choosing $t_B(0) = 0$ (so $\tau(0) = t_0$). 

It is important to realise that evaluating $R'(t,r)$ on the null cone 
$t =\hat{t}(r)$ is not the same as differentiating $\hat{R}(r) = 
R(\hat{t}(r),r)$ with respect to $r$. In fact, by evaluating 
(\ref{eq:R}) on the null cone, $\hat{R}$ is given by 
 \begin{equation} 
 \hat{R} = R(\hat{t}(r),r) = \left[{9 M(r) \over  2} \tau(r)^2 
 \right]^{1/3} 
 \label{eq:Rhat} 
 \end{equation} 
 which means that its derivative is given by 
 \begin{equation}          
 \frac{d \hat{R}}{dr} = \frac{d}{dr} [R(\hat{t}(r),r)]  = {3 \over 2 
R^2(\hat{t}(r),r)} [M'(r) \tau(r)^2 + 2 M(r)\tau(r) \tau'(r)]\,.      
 \end{equation}          
 Combining the above equation with the constraint (\ref{eq:condn}) 
gives a first order differential equation for $\hat{R}$. 
 \begin{equation} 
 \frac{d \hat{R}(r)}{d r} + {3 M(r) \tau(r) \over {\hat{R}(r)}^2} = 1 
\,.     \label{eq:ode} 
 \end{equation} 
 In summary, with our choice of coordinates we have recast the flat 
{\sc ltb} model in a form that allows us to locate the past null cone 
with ease. This has left us with one physical freedom to choose an 
arbitrary function of $r$. We could choose $\tau$ (or $M$) and 
substitute (\ref{eq:Rhat}) into (\ref{eq:ode}). Solution of this 
differential equation would determine $\hat{R}$ and thus any other 
quantity. If we instead decide to choose $\hat{R}$, that is, the area 
of the wavefront, then the model is trivially and fully specified by 
(\ref{eq:Rhat}) and (\ref{eq:ode}). It follows that 
 \begin{equation} 
 \tau(r) = {2 \hat{R}(r) \over 3} / \left(1 - \frac{d \hat{R}(r)}{d r} 
 \right) 
 \label{eq:tau2} 
 \end{equation} 
 and 
 \begin{equation} 
 M(r) = { \hat{R}(r) \over 2} \left(1 - \frac{d \hat{R}(r)}{d r} 
 \right)^2~~~~~(M(0) = 0) \,.   
 \label{eq:M2} 
 \end{equation} 
 To obtain the results of the next section, we will fix the geometry 
by choosing $\hat{R}$.  This will then determine $\tau(r)$ 
(equivalently $t_B(r)$) and $M(r)$ by the above two equations. The 
flat {\sc ltb} model will thus be fully specified in these coordinates 
and one could then propagate the data off the null cone by the 
comoving assumption.\\ The density on the null cone $\hat{\rho}(r)$ is 
found by evaluating (\ref{eq:rho}) on the null cone: 
 \begin{equation} 
 4 \pi \hat{\rho}(r) = 
 \frac{M'(r)}{{\hat{R}(r)}^2} 
 \label{eq:rhohat} 
 \end{equation} 
 and its  value at the origin depends on the time as characterised by 
the Hubble constant;\footnote 
 {Since measurements of the Hubble constant are taken at about $z < 
1$, we can take this to determine the age of the universe, $t_0$, at 
the central observer.} 
 \begin{equation} 
 \Omega = {4\pi \rho \over H^2} = 1 ~~\Rightarrow ~~ 
 \rho_0 = {H_0^2 \over 4\pi}, ~~t_0 = {2 \over 3 H_0}\,. 
 \end{equation} 

 \subsubsection{Redshifts} 
 It is of some importance that we state the relevant quantities in 
terms of redshifts. To do this, we use the fact that in the geometric 
optics limit, for two light rays emitted on the worldline at  $r_{em}$ 
with time interval $\delta t_{em} = t^{+}(r_{em}) - t^{-}(r_{em})$  
and observed on the central worldline with time interval $\delta 
t_{ob} =  t^{+}(0) - t^{-}(0)$ 
 \begin{equation} 
 1 + z = \frac{\delta t_{ob}}{\delta t_{em}}\,. 
 \label{eq:tint} 
 \end{equation} 
 The past radial null geodesics are given by 
 \[dt = - R'(t,r) dr \,,\] 
 so for an observer on a nearby worldline, the time interval changes 
by 
 \[d (\delta t) ~~=~~ dt^{+}-dt^{-} ~~=~~   \left[R'(t^{-},r) - 
R'(t^{+},r)\right] dr ~~=~~   - \frac{\partial}{\partial t} \left[ 
R'(t,r) 
 \right] \delta t dr \,. \] 
 Thus 
 \[d \ln \delta t   =   - \frac{\partial}{\partial t} \left[ R'(t,r) 
\right] dr \] 
 which means that the redshift, given by (\ref{eq:tint}), is 
 \begin{equation} 
 \ln (1 + z)   =   \int_{0}^{r_{em}} \dot{R}'(\hat{t},{r}) d {r} 
 \label{eq:lnz} 
 \end{equation} 
 where $\hat{t}(r)$ is the equation of the null cone.\footnote 
 {The standard formula $1 + z = (u^\mu k_\mu)_{em} / (u^\mu 
k_\mu)_{ob}$ is not useful in this gauge since $k^\mu = (R', -1, 0, 0) 
= (1, -1, 0, 0)$ is not geodesic, though it is tangent to the past 
null cone.} 
 To calculate $\dot{R}'(t,r)$ we differentiate (\ref{eq:R}) with 
respect to $t$ 
 \begin{equation} 
 \dot{R}' = {\dot{R} \over 3} \left[ {t_B' \over t - t_B} + {M' 
 \over M} \right], ~~~~~\dot{R} = \left[ {4 M \over 3 (t - t_B)} 
 \right]^{1/3} \,. 
 \end{equation} 
 Since $\hat{t}(r) = t_0 - r$ when we choose $R' = 1$ on the null 
cone, $\dot{R}'(\hat{t},r)$  is given by 
 \begin{equation} 
 \dot{R}' (\hat{t},r) = {1 \over 3} \left[ {4M \over 3 \tau} 
 \right]^{1/3} 
 \left[ {M' \over M} - {1 + \tau' \over \tau} \right] \,. 
 \end{equation} 
 After some manipulation of the above expression substituted into 
(\ref{eq:lnz}), we find that 
 \begin{equation} 
 \ln (1 + z) = \left({{4 M} \over {3 \tau}} \right)^{1/3}  - {1 \over 
3} 
 \int_{0}^{r_{em}} {\left({4 M} \over {3 \tau^4} \right)^{1/3}} dr \,. 
 \label{eq:z} 
 \end{equation} 
 Using (\ref{eq:tau2}) and (\ref{eq:M2}) this equation  may be written 
as 
 \begin{equation} 
 \ln (1 + z) =  \left(1 - \frac{d \hat{R}}{d r} \right)- 
 \frac{1}{2} \int_{0}^{r_{em}} \left( 1 - \frac{d \hat{R}}{d  {r}} 
\right)^{2} \,/\,{\hat{R}} \, d {r} 
 \label{eq:z2} 
 \end{equation} 
 so we can now determine the redshift-area distance relation. 

 \subsection{The Friedmann-Lema\^{\i}tre Limit} 
 The characterisation of the {\sc fl} limit is that the bang time 
surface is simultaneous. So $t_B(r) = {t_{B}}_{\mbox{\tiny FL}} = 
\mbox{constant}$; from whence  
 \begin{equation} 
 R_{\mbox{\tiny{{FL}}}}(t,r) = \left[ {9 \over 2} 
M_{\mbox{\tiny{{FL}}}}(r) (t - {t_{B}}_{\mbox{\tiny FL}}) ^2 
 \right]^{1/3},~~~~R_{\mbox{\tiny{{FL}}}}'(t,r) = {3 \over 2 
R_{\mbox{\tiny{{FL}}}}^2 } M_{\mbox{\tiny{{FL}}}}'(r) \,(t -
{t_{B}}_{\mbox{\tiny FL}})^{2} \,. 
 \label{eq:FLlim} 
 \end{equation} 
 The freedom left here in $M_{\mbox{\tiny{{FL}}}}(r)$ is just 
essentially the coordinate freedom, corresponding to the freedom of 
choice of $r $. The above relations determine the {\sc fl} density 
 \begin{equation} 
 \rho_{\mbox{\tiny{{FL}}}}(t) = \frac{1} {6 \pi (t - 
{t_{B}}_{\mbox{\tiny FL}})^2} 
 \end{equation} 
 which is spatially homogeneous as required, unaffected by 
$M_{\mbox{\tiny{{FL}}}}(r)$. It is usual to set ${t_{B}}_{\mbox{\tiny 
FL}} = 0$. We do not have a freedom to rescale the density by a 
constant because this is the critical density case. 

As we would eventually like to compare our {\sc ltb} model as chosen 
above to an underlying {\sc fl} model, it is appropriate to write the 
{\sc fl} limit in the same kind of coordinate system. Consider light 
rays coming in to the event ($t = t_1, r = 0$) in a {\sc fl} model. 
When we choose coordinates for which $R_{\mbox{\tiny{{FL}}}}'(t,r) = 
1$ on the null cone, the past null cone can be located by $\hat{t} = 
t_1 - r - {t_{B}}_{\mbox{\tiny FL}} = t_1 - r$. (We use $t_1$ rather 
than $t_0$ here, as we will need to distinguish {\sc ltb} and {\sc fl} 
values later on.)  As a limit of the flat {\sc ltb} model in these 
coordinates, the  {\sc fl} form of $M(r)$ is obtained from setting 
$\tau  = t_1 - r$ in (\ref{eq:condn}). This yields 
 \begin{eqnarray} 
 {M_{\mbox{\tiny{{FL}}}}}& = & {6} \left[ t_1^{1/3} - (t_1 - r)^{1/3} 
 \right]^3 ~~~~~~ ( M_{\mbox{\tiny{{FL}}}}(0) = 0 )  \label{MFL} \\ 
{{\hat{R}}_{\mbox{\tiny{{FL}}}}}& = & {3} \left[ t_1^{1/3} - (t_1 - 
r)^{1/3} 
 \right](t_1 - r)^{2/3} ~~~~~~ ( R_{\mbox{\tiny{{FL}}}}(0) = 0 ) \,. 
 \label{eq:Rrwhat} 
 \end{eqnarray} 
 We note that this in conjunction with (\ref{eq:FLlim}) implies that 
 \begin{equation} 
 R_{\mbox{\tiny{{FL}}}}(t,r) = 3 \left[ t_1^{1/3} - \left(t_1 - r 
 \right)^{1/3} \right] t^{2/3}~~~,~~~ R_{\mbox{\tiny{{FL}}}}' = 
\frac{3 {M_{\mbox{\tiny{{FL}}}}'}}{2 R_{\mbox{\tiny{{FL}}}}^{2}}  t^2 
= {t^{2/3} \over (t_1 - r)^{2/3}}\,. 
 \label{eq:r1} 
 \end{equation} 
 The {\sc rw} metric that results is, from (\ref{eq:r1}) and 
(\ref{eq:flatmet}), 
 \begin{equation} 
 ds^2 = - dt^2 + t^{4/3} \left\{ {1 \over (t_1-r)^{4/3}} dr^2 + {9} 
\left[ t_1^{1/3} - (t_1-r)^{1/3} \right]^2 d\Omega^2 
 \right\} \,.  \label{ds2FL} 
 \end{equation} 
 These coordinates are singular at the particle horizon, $r = t_1$ 
(when the past null cone of $t = t_1$ runs into the initial 
singularity). Thus they are valid for $0 \leq r < t_1$. The {\sc fl} 
redshift-distance formula can be obtained by inserting the {\sc fl} 
forms of $M(r)$ and $\tau(r)$ into equation (\ref{eq:z}). That is 
 \begin{equation} 
 z(r) = \left({t_1 \over t_1 - r} \right)^{2/3} - 1 
~~~~~~\Longleftrightarrow~~~~~~r(z) = t_1 \left[{(1 + z)^{3/2} - 1 
 \over (1 + z)^{3/2}} \right] \,. 
 \end{equation} 

 \subsection{Averaging and Fitting} 
 We want to compare and contrast total areas of wavefronts at given 
redshifts of an inhomogeneous model to that of the corresponding {\sc 
fl}  model of density equal to the inhomogeneous density perfectly 
smoothed. This must be done with respect to the inhomogeneous metric 
because physically the smoothing does not occur. 

Perhaps the crucial part of our analysis is that we ensure that we 
compare with the {\sc fl} model with the correct average density. We 
define the {\em average} or {\em background} {\sc fl} model to be the 
one that matches on at the particle horizon  where $\tau = 0$, $r = 
r_\Sigma$, using the Darmois-Israel boundary conditions 
\cite{bi:D27,bi:I66}. Matching first and second fundamental forms of 
this timelike (comoving) boundary surface $\Sigma$ gives 
 \begin{equation} 
 \left.{R_{\mbox{\tiny{LTB}}}}\right|_{\Sigma} = 
 \left.{R_{\mbox{\tiny{FL}}}}\right|_{\Sigma} 
 \end{equation} 
 and the background model must be parabolic if the inhomogeneous one 
is; or vice versa. 

The matching must hold over all of $\Sigma$; that is, at all times 
 --- so 
 \begin{equation} 
 \left.{{\dot{R}}_{\mbox{\tiny{LTB}}}}\right|_{\Sigma} = 
 \left.{{\dot{R}}_{\mbox{\tiny{FL}}}}\right|_{\Sigma} 
 \end{equation} and thus, by (\ref{eq:Rdot}), 
 \begin{equation} 
 \left.{M_{\mbox{\tiny{LTB}}}}\right|_{\Sigma} = 
 \left.{M_{\mbox{\tiny{FL}}}}\right|_{\Sigma}\,.  
 \end{equation} 
 Thus it is sufficient to match the masses at $\Sigma$, and 
synchronise the starting times (bang times) when 
$\left.{R_{\mbox{\tiny{LTB}}}}\right|_{\Sigma} = 0 = 
\left.{R_{\mbox{\tiny{FL}}}}\right|_{\Sigma}$. In general, we do not 
expect the {\sc fl} radial coordinate on $\Sigma$ 
($\left.{r_{\mbox{\tiny{FL}}}}\right|_{\Sigma}$) to be the same as the 
{\sc ltb} one there ($\left.{r_{\mbox{\tiny{LTB}}}}\right|_{\Sigma} = 
r_\Sigma$) since the coordinate condition $\hat{R'} = 1$ holds on the 
null cone, whose locus is model dependent. 

For a parabolic {\sc ltb} model with metric (\ref{eq:flatmet}) and 
density given by (\ref{eq:rho}), the background density 
$\rho_{\mbox{\tiny FL}}$ is the same as that obtained by integrating 
over constant time slices. 
 \begin{eqnarray} 
 \langle \rho \rangle_{t_0, r_{\Sigma}} &=& \left( {\int_0^{2\pi} 
\int_0^{\pi} 
 \int_0^{r_{\Sigma}} ~\rho \sqrt{{}^{3}g}~ dr d\theta d\Phi} 
\right)\,/\, 
 \left({\int_0^{2 \pi} \int_0^{\pi} \int_0^{r_{\Sigma}} 
~\sqrt{{}^{3}g}~ dr d\theta d\Phi }  \right) \nonumber \\ 
                                &=& \left({\int_{0}^{r_{\Sigma}} 
 \frac{M'}{R^{2} R'} \sqrt{{R'}^2 R^4} dr} \right)\,/\, \left({4\pi 
 \int_{0}^{r_{\Sigma}} \sqrt{{R'}^2 R^4}  dr } \right)  \nonumber \\ 
                                &=& \frac{3}{4 \pi} 
 \frac{M(r_{\Sigma})}{[R(r_{\Sigma},t_0)]^{3}}                 
 \nonumber \\ 
                                &=& \frac{1}{6 \pi (t_0 - 
t_{B}(r_{\Sigma}))^{2}} =  \rho_{\mbox{\tiny{FL}}} 
 \label{eq:tolavg} 
 \end{eqnarray} 
 where equation (\ref{eq:R}) was used. 

One important point that must be made here is that a covariant 
averaging procedure does not exist as yet. We have used here an 
averaging method which is `natural' for the comoving synchronous 
coordinates which lead to a $3 + 1$ foliation of spacetime. However, 
the same model in different (for example observational) coordinates 
would suggest a different averaging procedure which could conceivably 
yield different results. Therefore the claim \cite{bi:W76} that the 
wavefront areas obtained in the inhomogeneous model and the averaged 
model are the same already seems highly unlikely. 

 \section{Results} 
 \label{sec:results} 
 We use geometric units such that $G = c = 1$.  If we choose a unit of 
time $T_G$ {\it seconds} to be 1 geometric time unit ($gtu$), then the 
geometric units of length, mass, density, etc. are fixed by $ 1 \; glu 
= L_G = cT_G$ {\it metres}, $1 \; gmu = M_G = (c^3/G)T_G \; kg$, $1 \; 
gmu 
 \; glu^{-3} = \rho_G = (1/G)T_G^{-2} \; kg \; m^{-3}$.  For the 
purposes of this paper, we want units suitable to cosmological scales, 
so we specify that one cosmological time unit, $1 \; ctu$, is ten 
billion years 
 --- of the order of the age of the universe.  This gives us 
 \begin{center} 
 {\bf Cosmological Geometric Units} 
 \end{center} 
 \begin{center} 
 \begin{tabular}{|l|l|l|l|l|} 
 \hline 
              & Time & Length & Mass & Density \\ 
 \hline 
 Cosmological & $1 \; ctu$ 
                     & $1 \; clu$ 
                              & $1 \; cmu$ 
                                     & $1 \; cmu \; clu^{-3}$ \\ 
 \hline 
 SI           & $3.156 \times 10^{17} \; s$ 
                     & $9.461 \times 10^{25} \; m$ 
                              & $1.275 \times 10^{53} \; kg$ 
                                     & $1.505 \times 10^{-25} \; 
kg/m^3$ \\ 
 \hline 
 Astronomical & $10 \; Gyr$ 
                     & $3.066 \; Gpc$ 
                              & $6.409 \times 10^{22} \; M_\odot$ 
                                     & $1.505 \times 10^{-28} \; g/cc$ 
\\ 
 \hline 
 \end{tabular} 
 \end{center} 

The first subsection (\ref{sec:mod1}) gives a very simple model which 
satisfies the criteria for a reasonable cosmological model (with the 
classical Copernican principle dropped) and which provides a {\em 
proof that there exist physically reasonable density behaviours which 
lead to a nonzero magnification or shrinking}. It is obvious that 
averaging over the sky will not remove this effect since the model is 
already spherically symmetric. The second model (section 
\ref{sec:mod2}) does the same, but is smoother at the origin and 
displays interesting behaviour in redshift space, and the third model 
(section {\ref{sec:mod3}) is exactly homogeneous beyond a given 
comoving radius.

 \subsection{Form of Perturbation and General Results} 
 It is not easy to choose a form of area distance function for the 
inhomogeneous model which results in reasonable physical behaviour. So 
instead we choose it in  the form of a `perturbation' from a flat 
Friedmann model; that is, 
 \begin{equation} 
 \hat{R}(r)= \hat{R}^{u}_{\mbox{\tiny{{FL}}}}(r) (1 + \delta(r)) 
 \label{eq:hatR} 
 \end{equation} 
 where, from (\ref{eq:Rrwhat}), 
$\displaystyle{\hat{R}^{u}_{\mbox{\tiny{{FL}}}}(r) = 3[{t_{u}}^{1/3} - 
(t_{u} - r)^{1/3}]\,(t_{u} - r)^{2/3}}$ is the area function of an 
{\em underlying} {\sc fl} model of age $t_{1} = t_{u}$. (This 
`underlying' {\sc fl} model is  a mathematical device with no physical 
significance. It can not be considered a background or average model 
since we have not restricted $\delta(r)$ to average out to zero in any 
sense.) In principle, one should choose a density function and then 
determine the area distance function from it or risk the possibility 
of assuming  the result. However, if we can show that the above choice 
of $\hat{R}$ leads to a density profile with reasonable physical 
behaviour, this  would suffice 
 --- since if we had initially chosen that density function, it would 
lead to an $\hat{R}$ as chosen above. We will show that this is indeed 
the case and also indicate that the model is free of shell 
crossings.\footnote 
 {The necessary and sufficient conditions for there to be no shell 
crossings anywhere or at any time in the evolution of  a flat model 
with $R' > 0$  are that $M(r)$ be an increasing and $t_{B}(r)$   a 
decreasing function. They  were found (for all {\sc ltb} spacetimes) 
by Hellaby and Lake \cite{bi:HL85}.} 

Obviously, for $\delta(r)$ smooth and finite, $\hat{R}(r)$ is zero at 
the same places as $\displaystyle{\hat{R}^{u}_{\mbox{\tiny{FL}}}}(r)$, 
i.e. at $r=0$ and at $r = t_{u}$. For this form of perturbation, in 
terms of the convenient parametrisation $v = r/t_{u}$, we find 
 \begin{eqnarray} 
 X                &\equiv& \left[ \frac{1}{(1 - v)^{1/3}} - 1 \right] 
\\ 
 M                &  =   & \frac{3}{2} t_u X (1 - v) (1 + \delta) [2X 
(1 + \delta) - 
 \delta - 3 t_u X (1 - v) \delta']^2 \\ 
 \tau             &  =   & \frac{2 t_u X (1 - v) (1 + \delta)}{[2 X (1 
+ \delta) - 
 \delta - 3 t_u X (1 - v) \delta']} \\ 
 t_B              &  =   & t_0 - r - \tau \\ 
 8 \pi \hat{\rho} &  =   &\frac{2 X (1 + \delta) - \delta - 3 t_u X (1 
- v) 
 \delta'}{9[t_u X (1 - v) (1 + \delta)]^2} \{ 2 X (3 + 4\delta)(1 + 
\delta) -
 \delta (1 + \delta) - 3 t_u X (1 - v) (5 + 6 \delta) \delta'  
\nonumber \\ 
                 &+& 36 t_u X^2 (1 
 - v) (1 + \delta) \delta' - 9 {t_{u}}^2 X^2 (1 - v)^2[2 (1 + 
\delta)\delta'' 
 + \delta'^2]\} \\ 
 \frac{d \ln (1 + z)}{d z} & = &\{ 4 X (1 + 2 \delta) (1 + \delta) - 
\delta^2 
 - 6 t_u X (1 - v) (2 + 3 \delta) \delta' \nonumber \\ 
                 &+& 36 t_u X^2 (1 - v) (1 + \delta) 
 \delta' - 9 {t_u}^2 X^2 (1 - v)^2 [2 (1 + \delta) \delta'' + 
\delta'^2] \} 
 \,/\,[6 t_u X (1 - v) (1 + \delta)]\,. 
 \end{eqnarray} 
 If $\delta(0) \neq 0$ we find the unphysical limits $\tau(0) = 0$ and 
$\hat{\rho}(0) = \infty$. Thus we set $\delta(0) = 0$, obtaining the 
following limiting values: 
 \begin{eqnarray} 
 M(0) &=& \left. \frac{2 (1 - 3 t_u \delta'(r))^2 r^3}{9 {t_u}^2} 
 \right|_{r \rightarrow 0} = 0 \\ 
 \tau(0) &=& \frac{t_u}{1 - 3 t_u \delta'(0)} \label{eq:taumod} \\ 
 8 \pi \hat{\rho}(0) &=& \frac{4(1 - 3 t_u \delta'(0))^2}{3 {t_u}^2}\\ 
 \left. \frac{d \ln (1 + z)}{dr} \right|_{r = 0} &=& 
 \frac{2 (1 - 3 t_u \delta'(0))}{3 t_u} 
 \end{eqnarray} 
 and 
 \begin{eqnarray} 
 M(t_u) &=& 6 t_u (1 + \delta(t_u))^3 \label{eq:Mmod} \\ 
 \tau(t_u) &=& \left. (t_u - r) \right|_{r \rightarrow t_u} = 0 \\ 
 8 \pi \hat{\rho}(t_u) &=& \left. \frac{4 (3 + 4 \delta(r))} 
 {9 (t_u - r)^2} \right|_{r \rightarrow t_u} = \infty \\ 
 \left. \frac{d \ln (1 + z)}{d r} \right|_{t_u} &=& \left. \frac{2 (1 
+ 2 
 \delta(r))}{3 (t_u - r)} \right|_{r \rightarrow t_u} = \infty \,. 
 \end{eqnarray} 
 (The limits for the background {\sc fl} model are obtained by setting 
$\delta = 0 = \delta' = \delta''$ and replacing $t_u$ by $t_b$.)  From 
numerical experimentation we concluded, in order to avoid shell 
crossings, that $\delta(r)$ must remain sufficiently far away from 
zero over most if not the entire range of $r > 0$, and certainly near 
$r = t_u$. We want the proper time from the bang surface to the null 
surface on the central worldine to be the `true' age of the universe; 
that is, we want it to be $t_0$, the time at the origin of the {\sc 
ltb} model. By setting $\tau(0) = t_0$ in (\ref{eq:taumod}), the age 
of the underlying model is determined 
 \begin{equation} 
 t_{u} = t_0 \left( 1 - 3\, t_{1} \, \delta'(0) \right)\,. 
 \end{equation} 
 The parameter $t_{u}$ is the $r$-coordinate value at which the null 
cone of the {\sc ltb} model intersects the bang. We will average 
quantities on this scale; that is to say, we shall take $r_{\Sigma} = 
t_{u}$. We match this inhomogeneous universe to a flat {\sc fl} model 
at the surface $r_{\Sigma}$ by equating the masses and bang times at 
that point. This then determines the time $t_{1} = t_{b}$ in the {\em 
background} {\sc fl} model which we will use for our comparison. At $r 
= t_{u}$, $\hat{R} = 0$ and (\ref{eq:Mmod}) shows that at this point, 
 \[ M(t_{u}) = M^{u}_{\mbox{\tiny{{FL}}}}(t_{u}) \, (1 + 
\delta(t_{u}))^3 = 6 t_{u} \, (1 + \delta(t_{u}))^3 \,.\] 
 In the background {\sc fl} model the value of the mass at $\Sigma$ is 
$6 t_{b}$ and this is what we have to match the inhomogeneous mass to. 
This gives us a value of the  age for the background flat {\sc fl} 
model of 
 \begin{equation} 
 t_{b} = t_{u} \left(1 + \delta(t_{u}) \right)^{3} \,. 
 \end{equation} 

 \subsection{A Regular Model which Exhibits Shrinking and 
Magnification} 
 \label{sec:mod1} 
 The following simple example is physically well behaved, being free 
of shell crossings at all times in its evolution for $r \leq t_{u}$. 
Since $t_{B}' 
 \neq 0$ at the origin, the model is not as smooth there as one would 
like, but there are no physical problems. We choose $\delta(r)$ for 
our first model, LTB1, to be 
 \begin{equation} 
 \delta(r) = - \frac{1}{5} \sin \left({0.8 \pi r \over t_{u}}  
\right)\,. 
 \end{equation} When we set $\tau(0) = t_0 = 1$ then 
 \begin{equation} 
 t_{u} = t_0 \left( 1 + \frac{12 \pi}{25} \right) 
 \end{equation} 
 and the age for the background model FL1, after matching the masses, 
is 
 \begin{equation} 
 t_{b} = t_{u} \left(1 - \frac{\sin 0.8 \pi}{5} \right)^{3}\,. 
 \end{equation} 
 The calculation of the redshift was done by a numerical quadrature of  
(\ref{eq:z2}).  Figs. \ref{fig:rz1} and \ref{fig:rhoz1} compare the 
dependence of area distance and density on redshift, $\hat{R}(z)$ \& 
$\hat{\rho(z)}$, in LTB1 with the corresponding functions in the 
background model FL1. 

 \begin{figure} 
 \vspace*{0.3cm}
 \epsffile[0 0 399 241]{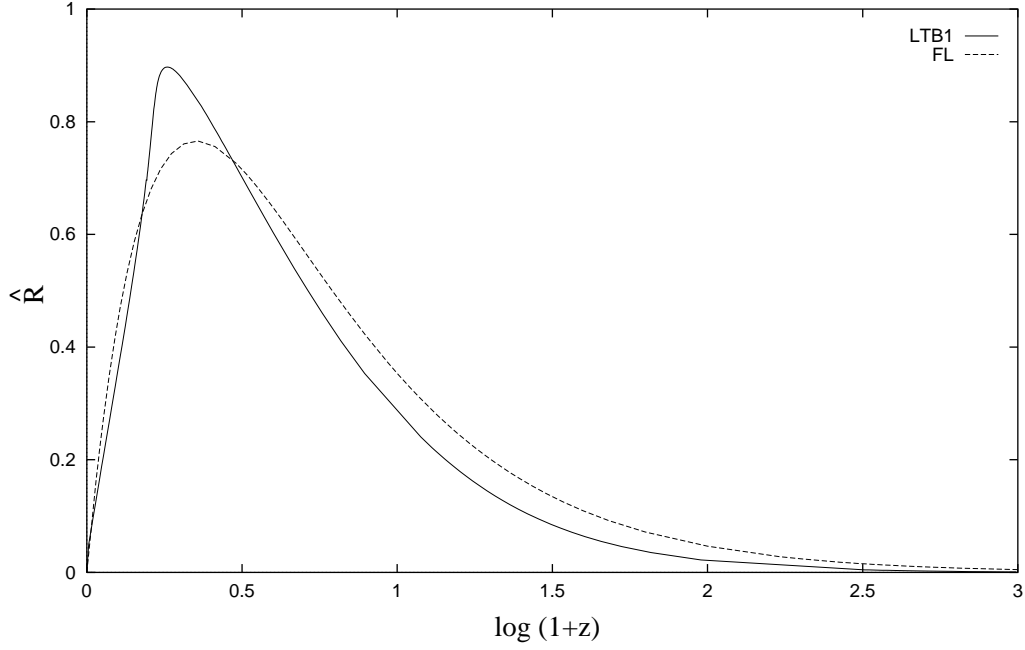} 
 \vspace*{0.5cm} 
 \caption{A plot of area distance against redshift $\hat{R}(z)$ \& 
$\hat{R}_{FL}(z)$ on the past null cone of the inhomogeneous model 
LTB1 and the corresponding background model FL1.  The units of $\hat{R}$ 
are cosmological length units, and all the figures use base 10 logs.  This 
shows that there are systematic shrinking ($\hat{R} > \hat{R}_{FL}$) and 
magnification ($\hat{R} < \hat{R}_{FL}$) effects due to purely radial 
lensing, which obviously cannot be removed by averaging over large angular 
scales or even the whole sky. Effects of true lensing in a more realistic 
universe would be imposed on top of this.} 
 \label{fig:rz1} 
 \end{figure} 
 \vspace*{.1cm}

 \begin{figure} 
 \vspace*{0.3cm} 
 \epsffile[0 0 396 239]{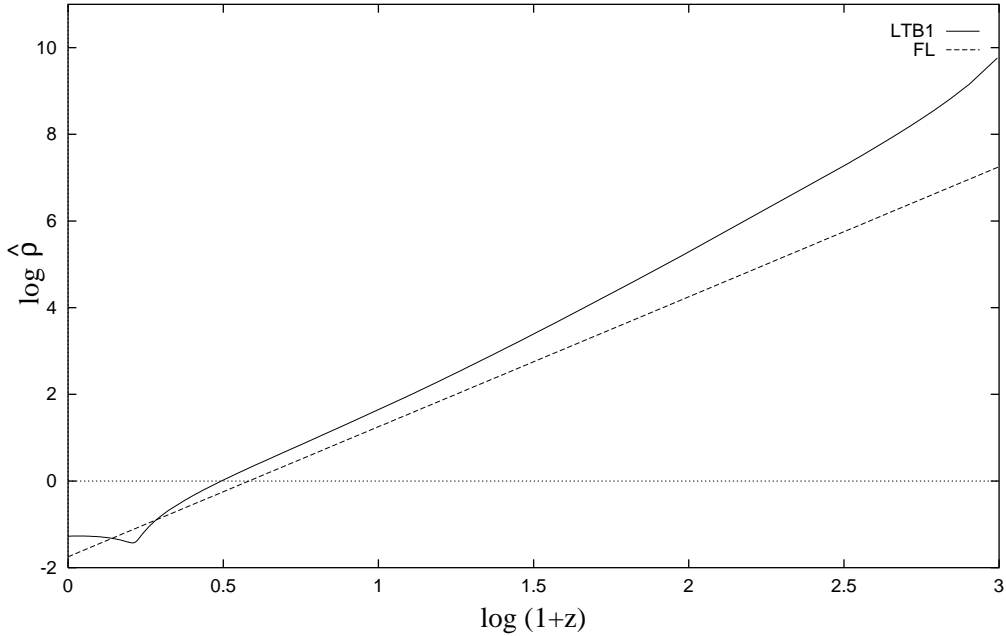} 
 \vspace*{0.5cm} 
 \caption{The density of matter $\hat{\rho}(z)$ \& 
$\hat{\rho}_{FL}(z)$ against redshift on the past null cone 
 in the models of {\sc fig.}~\protect{\ref{fig:rz1}}, LTB1 and its 
corresponding background model FL1.  The units are cosmological 
density units ($cmu \; clu^{-3}$).  When comparing with {\sc 
fig.}~\protect{\ref{fig:rz1}}, we see that roughly speaking, 
magnification occurs for obects in or just beyond an overdense region, 
and shrinking occurs for objects in or just beyond an underdensity.} 
 \label{fig:rhoz1} 
 \end{figure} 
 \vspace*{.1cm} 

It is important to plot these quantities in terms of the observable 
quantity $z$ for two  reasons. First of all, in the transformation $r 
\rightarrow z$, the possibility exists that the area distances of the 
flat and inhomogeneous models might transform into each other. 
Secondly, under certain circumstances the redshift becomes disordered 
with distance and unexpected behaviour might occur, as the  following 
model illustrates. 

 \subsection{A Regular Model with Multivalued Observations} 
 \label{sec:mod2} 
 As an illustration of how different the physical quantities plotted 
against radial coordinate $r$ as opposed to those same quantities 
plotted against redshift $z$ may appear, we present here an {\sc ltb} 
model for which the redshift becomes disordered with distance at some 
points and then ordered again at later points. 

The universe is chosen as above but with a `perturbation function' of 
 \begin{equation} 
 \delta (r) = -{1\over4}\sin \left(\frac{0.75 \pi r}{t_{u}} \right) -
 \frac{1}{4^{6}} \left(\frac{11}{ \pi} + \frac{3}{2} 
 \right) \left[ 1 - \cos \left(\frac{4 \pi r}{t_{u}} \right)\right]\,. 
 \end{equation} 
 This model, which we call LTB2,  is also free of shell crossings at 
any time for $r \leq t_{u}$ and has a completely smooth and regular 
origin (where $t_{B}' = 0$). Setting $\tau(0) = t_0 = 1$ once again 
gives 
 \begin{equation} 
 t_{u} = t_0 \left( 1 + \frac{9 \pi}{16} \right) 
 \end{equation} 
 and the appropriate background model FL2 has age 
 \begin{equation} 
 t_{b} = t_{u} \left(1 - \frac{\sin 0.75 \pi}{4} \right)^{3}\,. 
 \end{equation} 
 which are different from those of LTB1.  Figs. 
 \ref{fig:rr2}-\ref{fig:rhoz2} show $\hat{R}$ and $\hat{\rho}$ against 
$r$ and against $z$ for LTB2 and FL2. 

 \begin{figure} 
 \vspace*{0.5cm} 
 \epsffile[0 0 401 239]{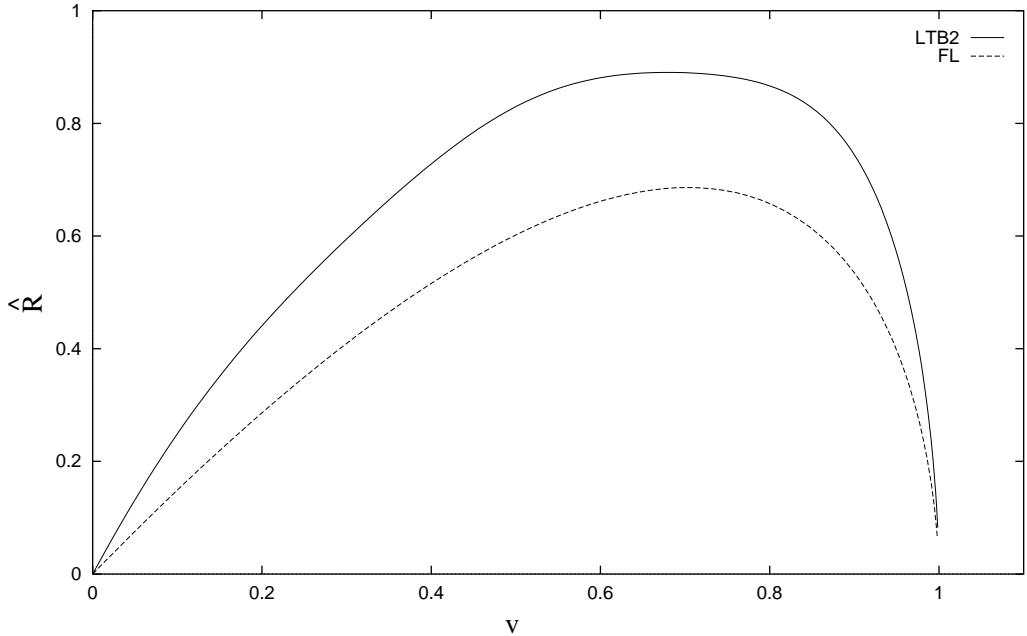} 
 \vspace*{0.5cm} 
 \caption{The area distance functions $\hat{R}(r)$ \& 
${\hat{R}}_{FL}(r)$ on the past null cones of the second model, LTB2, 
and its corresponding background model, FL2, given in cosmological 
units.   The horizontal variable is $v = r/t_u$ or $r/t_b$ for LTB2 
and FL2 respectively.  The physical behaviour of LTB2 as plotted 
against $v$ appears normal.} 
 \label{fig:rr2} 
 \end{figure} 
 \vspace*{.2cm} 

 \begin{figure} 
 \vspace*{0.5cm} 
 \epsffile[0 0 396 238]{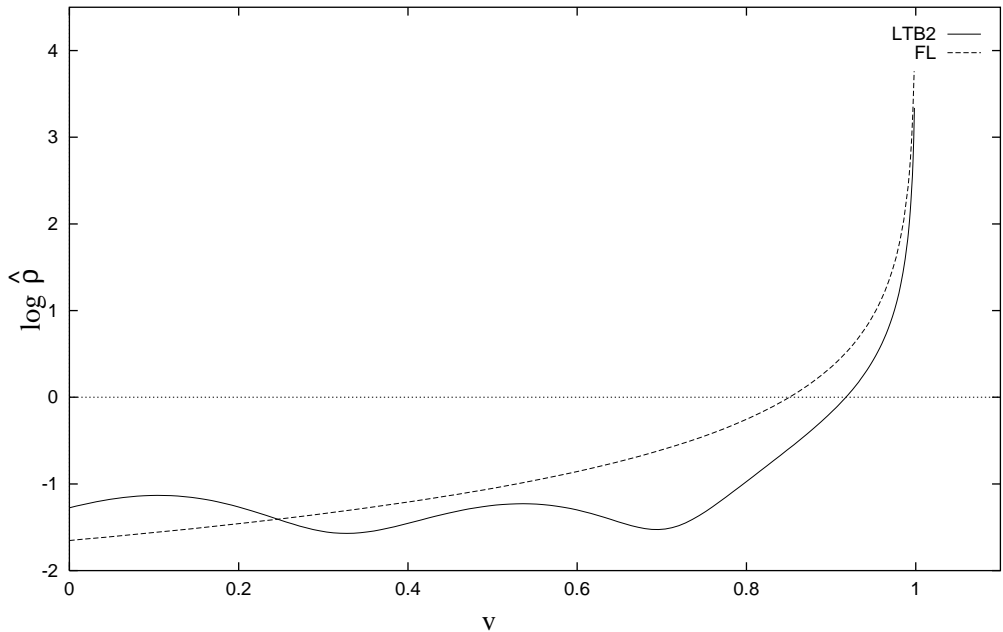} 
 \vspace*{0.5cm} 
 \caption{The densities on the past null cone for LTB2 and FL2, 
$\hat{\rho}(r)$ and ${\hat{\rho}}_{FL}(r)$, in cosmological units.  
Again, LTB2's inhomogeneous profile vs $v$ appears quite acceptable.} 
 \label{fig:rhor2} 
 \end{figure} 
 \vspace*{.2cm}

 \begin{figure} 
 \vspace*{0.5cm} 
 \epsffile[0 0 401 243]{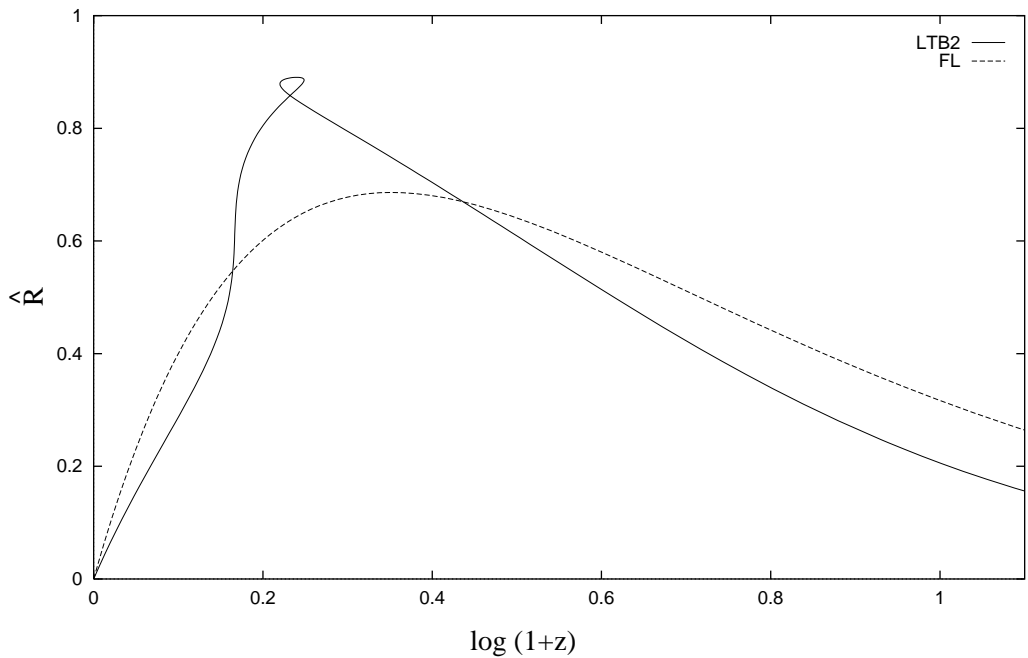} 
 \vspace*{0.5cm} 
 \caption{Area distance against redshift $\hat{R}(z)$ \& 
${\hat{R}}_{FL}(z)$ for LTB2 and FL2. The interesting point to note is 
that at some redshifts the area distance in LTB2 is multivalued.  For 
$\log_{10} (1 + z) \geq 1$ the graph looks very much like {\sc 
fig.}~\protect{\ref{fig:rz1}}.} 
 \label{fig:rz2} 
 \end{figure} 
 \vspace*{.2cm}

 \begin{figure} 
 \vspace*{0.5cm} 
 \epsffile[0 0 397 243]{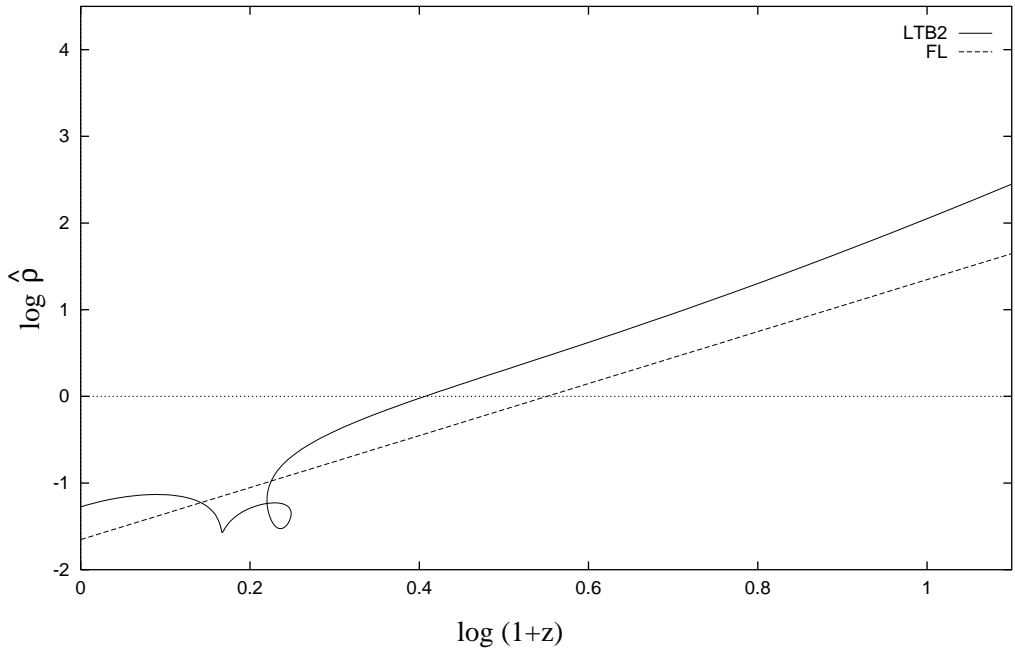} 
 \vspace*{0.5cm} 
 \caption{The densities, $\hat{\rho}(z)$ and ${\hat{\rho}}_{FL}(z)$, 
against redshift for LTB2 and FL2.  Note the quaint `looping' 
behaviour.  For $\log_{10} (1 + z) \geq 1$ the graph looks very much 
like {\sc fig.}~\protect{\ref{fig:rhoz1}}.} 
 \label{fig:rhoz2} 
 \end{figure} 
 \vspace*{.2cm} 

This model provides a good illustration of why one has to be careful 
in ascribing physical behaviour in a certain coordinate system. Viewed 
as functions of $r$, $\hat{R}$ and $\hat{\rho}$ have fairly standard 
behaviour, but viewed in terms of the observable quantity $z$, the 
density and area distance become multivalued. Hence, three objects 
with the same intrinsic luminosity located at different distances 
appear at the same $z$, with three different apparent luminosities (or 
area distances). 

Our numerical experiments indicate that the redshift on the light cone 
is most sensitive to perturbations in the vicinity of the maximum in 
$\hat{R}(z)$. All our models in which $dz\,/\,dr$ became negative did 
so in this region. The looping behaviour in the $\hat{R}$ vs $\log (1 
+ z)$ plot occurs when the maximum and minimum in the $\log (1 + z)$ 
vs $r$ graph bracket the maximum in the $\hat{R}$ vs $r$ graph. 
Similarly, perturbations more easily generate a maximum and minimum in 
the $\hat{\rho}(z)$ near the maximum in $\hat{R}$, hence the loop in 
that graph. 

 \subsection{A Swiss-Cheese Type Model} 
 \label{sec:mod3}
 In order to emphasise the effect of inhomogeneity on the past null 
cone, we consider a model which is homogeneous on the large scale.  We 
construct a local inhomogeneity which is properly matched onto an exact 
{\sc fl} model at some comoving radius, $r = r_j$, somewhere between us 
and the particle horizon.  In other words, the outer regions are exactly 
homogeneous.  This is a simple model of a universe which is {\sc FL} on 
some averaging scale less than the horizon size, and is seen by an 
observer located in an overdensity. 

In particular, for LTB3, we set $r_j = 0.85 t_u = 2.352$, we retain 
the inhomogeneous LTB2 model for the interior, $r \leq r_j$, and for 
the exterior, $r \geq r_j$, we use the matching {\sc fl} model, FLJ, 
with the same mass and bang time as LTB2 at the boundary 
$r_j$.\footnote 
 {In order to make the radial coordinate continuous at this matching, 
we use $\hat{R}_{FLJ}=3[t_3^{1/3}-(t_2-r)^{1/3}](t_2-r)^{2/3}$, 
$M_{FLJ}=6[t_3^{1/3}-(t_2-r)^{1/3}]^3$, where 
$t_2=\tau_{LTB}(r_j)+r_j$ and 
$t_3^{1/3}=\tau_{LTB}^{1/3}(r_j)+(M_{LTB}(r_j)/6)^{1/3}$.  These 
differ from the purely {\sc fl} expressions because our definition of 
$r$ is also path dependent.  It turns out that $t_b=t_3$.} 
 Obviously the background model FL3 that we fit at the particle 
horizon is identically the one matched at $r_j$, FLJ, but written in 
the form 
 (\ref{MFL})-(\ref{ds2FL}) 
 --- i.e. with a different $r$ coordinate. 

 \begin{figure} 
 \vspace*{-0.1cm} 
 \epsffile[0 0 401 243]{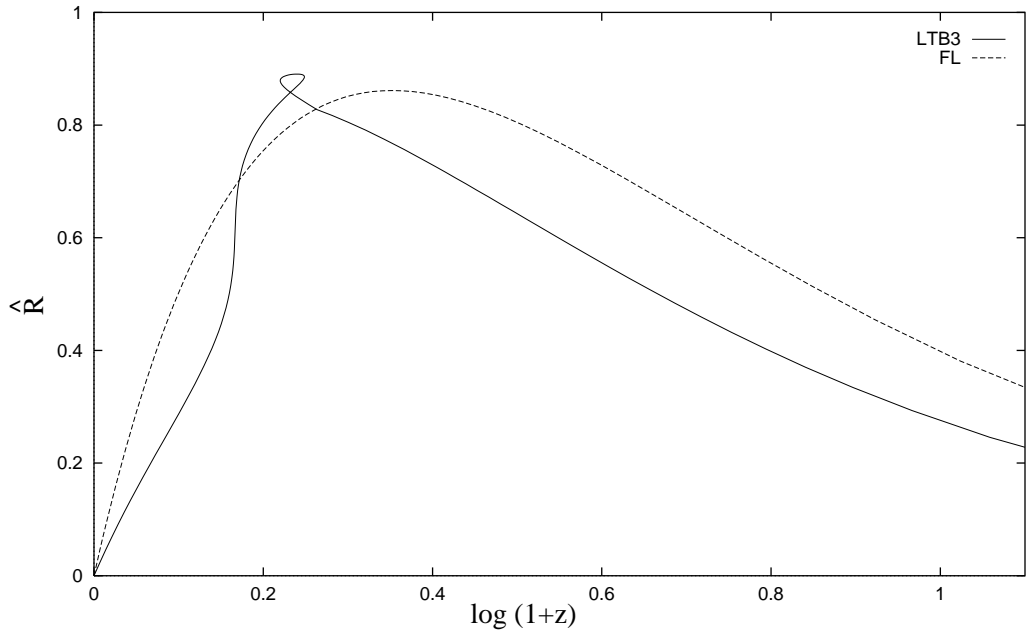} 
 \vspace*{0.3cm} 
 \caption{Area distance against redshift $\hat{R}(z)$ \& 
$\hat{R}_{FL}(z)$ for the swiss-cheese model LTB3 and its 
corresponding background model FL3.  The boundary is located at 
$\log_{10}(1+z_j)=0.2635$, with the model being identical to LTB2 for 
smaller $z$, and exactly {\sc fl} for larger $z$.  Though the 
background and matched {\sc fl} models are identical, the $\hat{R}(z)$ 
curves do not match up in the exterior region 
 --- there is a constant displacement parallel to the $\log(1+z)$ 
axis.  This emphasises the fact that the $z$ integral is path 
dependent, and so the value of $z$ at any given area distance is 
affected by the warping of the null cone in the inhomogeneous 
interior.} 
 \label{fig:rz3} 
 \end{figure} 
 \vspace*{2mm}

 \begin{figure} 
 \vspace*{-0.1cm} 
 \epsffile[0 0 397 243]{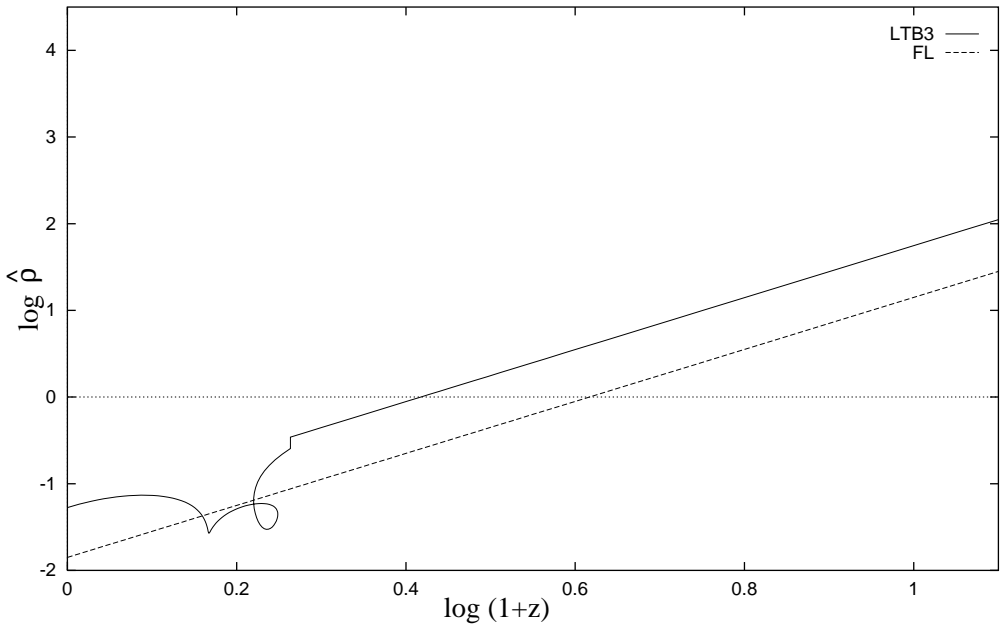} 
 \vspace*{0.3cm} 
 \caption{Density against redshift $\hat{\rho}(z)$ \& 
$\hat{\rho}_{FL}(z)$ for the swiss-cheese model LTB3 and its 
background model FL3.  The density jumps at the boundary, 
$\log_{10}(1+z_j)=0.2635$, but this is not unrealistic.  (Within the 
constraints of a parabolic model, it could only have been made 
continuous at the expense of creating shell crossings.)  As with the 
area distance, the two $\hat{\rho}(z)$ curves do not co-incide in the 
{\sc fl} exterior, because the shape of the null cone is determined by 
the spacetime geometry it encounters along its path.  The boundary 
could have been put at any $z$ value, and it is evident from  {\sc 
figs.}~\protect{\ref{fig:rz2}} and \protect{\ref{fig:rhoz2}} 
especially that in general the standard {\sc fl} relations are not 
recovered in the {\sc fl} exterior. 
 } 
 \label{fig:rhoz3} 
 \end{figure} 
 \vspace*{2mm}

Figs \ref{fig:rz3} and \ref{fig:rhoz3} show clearly that the 
 null cone 
 relations $\hat{R}(z)$ and $\hat{\rho}(z)$ in this model (LTB3) do 
not return to the standard {\sc fl} ones in the {\sc fl} region beyond 
the boundary, $z > z_j$.  This is because the redshift $z$ is path 
dependent and the inhomogeneous interior distorts the null cone.  
After passing through the inhomogeneity, the null cone, and thus the 
value of $\log_{10}(1+z)$, are displaced relative to a model which is 
{\sc fl} throughout.  This holds generally for any choice of 
inhomogeneous interior and boundary position. 

 \section{Conclusions} 
 \label{sec:conc} 
 The general belief that photon conservation implies that the total 
area of an incoming wavefront must be the same as in the background, 
matter-averaged, {\sc fl} model is contradicted by our results. 
 The functions $\hat{R}(z)$ and $\hat{\rho}(z)$ in an inhomogeneous 
model differ from their standard {\sc fl} forms, because the redshift 
$z$ is path dependent and the null cone is warped.  Thus, even if the 
inhomogeneous model is exactly homogeneous beyond some radius, the 
standard {\sc fl} forms are not recovered.  This means that a universe 
which is {\sc fl} {\em on average} will not in general present standard 
{\sc fl} forms of $\hat{R}(z)$ and $\hat{\rho}(z)$ to an observer. 

 The spherically symmetric model used here is simple but effective, 
since averaging over direction cannot change the results.  In more 
realistic models of the lumpy universe this effect will still be 
present, and we expect full gravitational lensing to occur, resulting 
in more significant deviations from the {\sc fl} formula. 

  This investigation used a parabolic {\sc ltb} model, where the areal 
radius $\hat{R}$ is also the area distance of the 2-sphere wavefronts 
of the past null cone. The density in the {\sc ltb} model is averaged 
to give a background Einstein-de Sitter ($\Omega = 1$) model, and it 
is tested against this model. Although there exists no covariant way 
to perform this averaging, we use the `natural' one defined by the use 
of junction conditions, here equivalent to the one used in 
astrophysical problems: that is, averaging by integrating the rest 
mass and proper volume on constant time slices.  
 More importantly, the functions $\hat{R}(z)$ and $\hat{\rho}(z)$ in 
the examples are obviously perturbed away from the standard {\sc fl} 
ones, so that no one {\sc fl} model can give the same $\hat{R}$ and 
$\hat{\rho}$ values as an inhomogeneous one at a variety of $z$ 
values. 

 The results show that it is quite easy to have areas in the 
inhomogeneous models  which differ significantly from areas in the 
background, matter-averaged {\sc fl} model. The result may either be 
shrinking (the background {\sc fl} area distance underestimates the 
real area distance at that redshift) or magnification (background {\sc 
fl} area distance is an overestimate). The presence of loops in the 
$\hat{R}$-$z$ and $\hat{\rho}$-$z$ graphs is analogous to the well 
known `finger of God' effect familiar in redshift maps of the galaxy 
distribution. 

  Our results and conclusions generally agree with those of 
 Kurki-Suonio and Liang \cite{bi:K-SL92}, who calculated obervational 
relations numerically in 4 hyperbolic {\sc ltb} models out to $z = 
0.5$.  They generated mild and strong deviations from the {\sc fl} 
observational relations, with `bang time' inhomogeneities and `areal 
radius' inhomogeneities having opposite effects on the oberved matter 
distribution in redshift space, compared with the present day 
distribution.  Their method did not permit the redshift to be 
disordered with distance. 

  Whilst the major aim of this paper has been the above thesis, the 
choice of radial coordinate which locates the observer's null cone 
will be of use in future analyses of observations in these isotropic 
dust models. 

  An important caveat is that since the {\sc ltb} model does not allow 
for formation of caustics in the null cone of the central observer, it 
cannot be considered a useful model for obtaining quantitative `real 
world' results.  Rather this paper should be viewed as a proof that 
even purely radial lensing distorts the area distance-redshift 
relation significantly. If the observer moves away from the central 
position, then continuity ensures that the radial effects found here 
will still be present, and the effects of true lensing will be 
superimposed. As argued in \cite{bi:EBD96,bi:SEB96}, we expect 
caustics to skew the area towards larger values, so that most objects 
in the universe are demagnified. 

  The importance of all this is that it opens up the way for 
considering the effects of lensing by inhomogeneities on large 
angular-scale number counts and 
 area distances 
 as opposed to limiting discussion to lensing effects on small scales.

 \end{document}